\pdfoutput=1
\documentclass[journal]{IEEEtran}
\IEEEoverridecommandlockouts
\usepackage{color}
\usepackage{subfigure}
\usepackage[ruled,vlined,algo2e]{algorithm2e}
\usepackage{algorithm}
\usepackage{algorithmic}
\usepackage{graphicx}
\usepackage{amssymb}
\usepackage{amsmath}
\usepackage[numbers,sort&compress]{natbib}
\usepackage{booktabs}
\usepackage{multirow}

\usepackage[hidelinks]{hyperref}

\begin{document}

\title{Wireless Power Control Based on Large Language Models}

\author{Jiacheng Wang,~\IEEEmembership{Graduate Student Member,~IEEE,}
        Yucheng Sheng,~\IEEEmembership{Graduate Student Member,~IEEE,}
        \\Le Liang,~\IEEEmembership{Member,~IEEE,}
        Hao Ye,~\IEEEmembership{Member,~IEEE,}
        and Shi Jin,~\IEEEmembership{Fellow,~IEEE}
\thanks{Jiacheng Wang, Yucheng Sheng, Le Liang and Shi Jin are with the School of Information Science and Engineering, Southeast University, Nanjing 210096, China (e-mail: wangjiacheng@seu.edu.cn; shengyucheng@seu.edu.cn; lliang@seu.edu.cn; jinshi@seu.edu.cn).}
\thanks{Hao Ye is with the Department of Electrical and Computer Engineering, University of California, Santa Cruz, CA 95064, USA (e-mail: yehao@ucsc.edu).}
}

\maketitle

\begin{abstract}
This paper investigates the power control problem in wireless networks by repurposing pre-trained large language models (LLMs) as relational reasoning backbones. In hyper-connected interference environments, traditional optimization methods face high computational cost, while standard message passing neural networks suffer from aggregation bottlenecks that can obscure critical high-interference structures. In response, we propose PC-LLM, a physics-informed framework that augments a pre-trained LLM with an interference-aware attention bias. The proposed bias tuning mechanism injects the physical channel gain matrix directly into the self-attention scores, enabling explicit fusion of wireless topology with pre-trained relational priors without retraining the backbone from scratch. Extensive experiments demonstrate that PC-LLM consistently outperforms both traditional optimization methods and state-of-the-art graph neural network baselines, while exhibiting exceptional zero-shot generalization to unseen environments. We further observe that topology-relevant relational reasoning is concentrated in shallow layers, whereas deeper layers encode task-irrelevant semantic noise. Motivated by this finding, we develop a lightweight adaptation strategy that reduces model depth by 50\%, significantly lowering inference cost while preserving state-of-the-art spectral efficiency.
\end{abstract}


\IEEEpeerreviewmaketitle

\section{Introduction}\label{sec:intro}

The evolution toward sixth-generation (6G) wireless ecosystems envisions ubiquitous connectivity characterized by ultra-reliability and massive capacity. To support the heterogeneous spectrum of nodes—encompassing autonomous vehicles, industrial sensors, and extended reality (XR) hardware—ultra-dense network architectures have become indispensable. 
In particular, aggressive spatial reuse mechanisms, which allow multiple transmitter-receiver pairs to concurrently share the same frequency band, have emerged as a crucial strategy for maximizing spectral efficiency (SE) in such dense deployments.
However, this inherent reliance on spatial reuse also creates severe interference. The resultant hyper-connected environments are plagued by complex, mutual interference patterns that scale non-linearly with network density, making contention for limited spectral resources a major bottleneck. 

Resource sharing requires judicious allocation for mitigating interference and optimizing resource utilization in ultra-dense wireless networks. However, traditional resource allocation designs fall short in massive connectivity scenarios, where strictly coupled interference constraints pose a severe obstacle to deriving optimal solutions. In this context, mathematical optimization frameworks grounded in information theory and game theory have been extensively explored. Nevertheless, the objective of maximizing sum rates renders the power control problem non-convex and NP-hard. While classical iterative algorithms such as WMMSE \cite{christensen2008wmmse} and FPLinQ \cite{shen2018fp1, shen2018fp2} can theoretically attain stationary points, they are plagued by prohibitive computational complexity and, more critically, numerical instability under stringent fairness constraints (e.g., harmonic utility). Consequently, reliance on such computationally intensive iterations becomes intractable for real-time applications as network density escalates.

As an effective tool to address problems of high dimensionality, deep learning (DL) has long been exploited for resource allocation design in wireless networks \cite{qin2024ai}. Multi-layer perceptrons (MLPs) \cite{sun2018learning}, convolutional neural networks (CNNs) \cite{cui2019spatial}, and deep reinforcement learning (DRL) frameworks \cite{liang2019spectrum} have been investigated to approximate optimization policies. More recently, graph neural networks (GNNs) have been further studied in \cite{shen2020graph, shen2023graph, naderializadeh2022state, naderializadeh2023learning} to exploit the permutation-equivariant property of wireless interference graphs. Similar system setups have been further considered in \cite{dai2024survey}, where GNNs are utilized to enable efficient resource allocation. Despite these advances, many existing GNNs follow the message passing neural network (MPNN) paradigm, which inherently relies on node-centric aggregation. In ultra-dense networks, where interference strengths can vary by several orders of magnitude, such aggregation creates a severe information bottleneck \cite{alon2021bottleneck}. Specifically, by indiscriminately compressing diverse channel state information (CSI) from all interacting user pairs into fixed-size node embeddings, MPNNs function as lossy compressors. Consequently, they suffer from signal dilution, where critical strong interference structures are numerically drowned out by the accumulation of diverse weak interference. This lossy compression imposes a performance ceiling, making it challenging for MPNN-based methods to consistently surpass traditional iterative benchmarks such as WMMSE \cite{christensen2008wmmse}.

In light of these limitations, the Transformer architecture \cite{vaswani2017attention} emerges as a compelling alternative to bypass the aggregation bottleneck. Through global self-attention, Transformers can directly model all-to-all interactions and capture complex relational dependencies without repeated and lossy local aggregation. Building upon this architectural advantage, large language models (LLMs) offer a further leap. Having demonstrated unprecedented representational capabilities across language \cite{devlin2019bert, brown2020language} and vision \cite{dosovitskiy2021image} tasks, LLMs possess rich structural priors. We hypothesize that the relational structures learned during LLM pre-training can be repurposed for wireless interference modeling, since both language and wireless networks involve structured interactions among many entities, albeit in different modalities. Specifically, the ability of LLMs to capture complex, long-range contextual dependencies structurally mirrors the challenge of modeling non-local, mutual interference coupling. This perspective suggests that pre-trained LLMs may serve as general relational reasoning backbones for physical systems when appropriately adapted. 
Recent studies in wireless communications have begun exploring LLM-based pipelines. The key obstacle, however, lies in the modality mismatch between continuous-valued wireless signals and the tokenized input of standard Transformer architectures. To mitigate this mismatch, several adaptation strategies have been proposed.
For instance, a beam prediction framework has been proposed in \cite{sheng2025beam}, which adopts a hybrid input strategy to activate the model's reasoning capabilities. Similar strategies have been further considered in \cite{Liu2024channel}, where learned CSI representations are embedded directly into the model to treat signal features as intrinsic inputs.

Despite these advancements, the potential of LLMs for medium access control (MAC) layer optimization remains largely unexplored. A key reason is that standard LLM architectures lack a native mechanism to effectively ingest interference topology. Unlike point-to-point physical-layer tasks, which primarily focus on signal reconstruction, MAC layer problems involve complex decision-making among interacting users, where the global interference topology is paramount. This creates a critical structural incompatibility: Standard Transformer attention is inherently feature-driven, relying on the dot-product similarity of latent embeddings. In wireless settings, however, the most important relationships are not necessarily between nodes with similar feature embeddings, but between nodes with strong physical interference coupling. Consequently, directly applying standard Transformer architectures to high-dimensional interference graphs can lead to severe information loss and sub-optimal resource orchestration.

To address this challenge and better harness the reasoning capabilities of LLMs for MAC layer optimization, we propose an interference-aware bias tuning mechanism that transforms a pre-trained LLM into a physics-informed graph Transformer. Drawing inspiration from AlphaFold2 \cite{jumper2021highly}, which successfully encoded 3D geometric constraints into attention mechanisms via pair biases, we adapt this paradigm to the wireless domain. Specifically, instead of relying solely on conventional feature embeddings, we inject the channel gain matrix directly into the self-attention module as an explicit physical bias term. This integration fundamentally reconfigures the attention landscape, compelling the model to prioritize nodes based on physical interference intensity rather than feature similarity. By selectively activating and fine-tuning the model's shallow structural knowledge via low-rank adaptation (LoRA) \cite{hu2021lora}, this approach efficiently strips away irrelevant semantic layers while retaining the Transformer's potent global reasoning capabilities. The main contributions of this paper are summarized as follows:
\begin{itemize}
    \item We propose PC-LLM, a novel framework that repurposes pre-trained LLMs for MAC-layer power control. To overcome the topological insensitivity of standard Transformers, we design an interference-aware bias tuning mechanism that injects the channel gain matrix directly into the attention mechanism via a lightweight projector. This enables an efficient attention-level fusion of physical topology and pre-trained structural priors, bypassing the aggregation bottleneck of MPNNs.
    \item Extensive experiments demonstrate that PC-LLM consistently outperforms the optimal WMMSE benchmark and state-of-the-art GNN baselines across a wide range of network conditions. We further observe favorable scaling behavior regarding data heterogeneity, where learning from diverse scenarios improves generalization compared with models trained on a single distribution. Furthermore, PC-LLM possesses exceptional zero-shot robustness, maintaining high spectral efficiency even when extrapolated to unseen interference distributions without additional fine-tuning.
    \item We provide an analysis of the layer-wise modality alignment between LLMs and wireless physics. We uncover an interesting phenomenon: The relational reasoning required for interference management is concentrated in shallow layers, while deeper layers encode task-irrelevant linguistic semantics. This insight establishes a guideline for minimalist adaptation, allowing us to truncate the model significantly (e.g., by 50\%) to reduce inference latency while preserving state-of-the-art performance.
\end{itemize}

The rest of the paper is organized as follows. In Section II, we provide a comprehensive review of the related literature regarding GNN-based resource management and the application of LLMs in wireless communications. Section III introduces the system model for wireless interference networks and formulates the power control optimization problem. Then, the proposed PC-LLM framework is presented in Section IV, where we detail the interference-aware bias tuning mechanism and the efficient training strategy. Simulation results are provided in Section V to validate the performance and generalization of our approach. Finally, the conclusion is drawn in Section VI.

\section{Related Work}\label{sec:related_works}
\subsection{GNN-based Resource Management}\label{subsec:related_works_GNN}
Owing to their inherent capability to capture topological structures and desirable scalability, GNNs have been extensively leveraged as the foundational architecture for next-generation network modeling. Theoretical frameworks have been established in \cite{shen2020graph}, where resource management problems are formulated as graph optimization tasks underpinned by the universal permutation equivariance property. Building upon this theoretical basis, GNNs efficiently exploit spatial dependencies to facilitate scalable solutions. For combinatorial problems such as link scheduling, graph embedding techniques have been investigated in \cite{lee2021graph} to maximize network throughput while significantly reducing the requisite training samples. More sophisticated architectures, such as GRLinQ \cite{shan2025grlinq} which integrates an MPNN with reinforcement learning, have demonstrated enhanced generalization capabilities and facilitated distributed execution.
To bridge the gap between theoretical performance and practical deployment, recent efforts have focused on addressing signaling overhead and strict constraint satisfaction. To circumvent the dependency on full global CSI, a sparse graph isomorphism network is proposed in \cite{wang2023sparse}, utilizing an edge-sparsifying mechanism to report only critical CSI features. Simultaneously, to rigorously enforce constraints within learning-based frameworks, GNNs have been effectively integrated with primal-dual methods in \cite{naderializadeh2022state, naderializadeh2023learning, liang2023primal}.Beyond static configurations, the adaptability of resource management policies to dynamic network has been improved through the integration of meta-learning and GNNs \cite{zhao2024meta, huang2025meta}. Specifically, a modular meta-learning approach is explored in \cite{nikoloska2023modular} to orchestrate power control in networks with arbitrarily time-varying topologies. For scenarios characterized by temporal shifts in CSI distribution, a meta-gating framework has been developed in \cite{hou2023meta}, which dynamically weighs the importance of historical periods to ensure fast and continuous resource optimization.

Notwithstanding these strides, underlying standard MPNNs impose an inherent representational bottleneck, which often limits their ability to consistently surpass robust benchmarks like WMMSE \cite{christensen2008wmmse} and FPLinQ \cite{shen2018fp1}.
We attribute this performance ceiling to the their node-centric message aggregation mechanism. While pairwise messages preserve local interaction details, the subsequent aggregation irreversibly squashes these diverse signals into a single fixed-size vector, causing severe signal dilution that obscures critical interference topologies.
The graph Transformer paradigm \cite{dwivedi2021generalization, ying2021transformers}—exemplified by the success of AlphaFold2 \cite{jumper2021highly} in biology—emerges as a compelling alternative to bypass this aggregation bottleneck. By maintaining a global pairwise attention matrix, it explicitly resolves all-to-all interactions before feature fusion, effectively avoiding the premature compression of MPNNs.
However, training such high-capacity models from scratch entails prohibitive sample complexity. Furthermore, standard attention modules lack the native ability to encode continuous interference topology. These challenges motivate our pivot toward adapting pre-trained Transformer backbones, leveraging their established relational priors to efficiently model the interference manifold without the cost of ground-up training.

\subsection{LLMs for Wireless Communication}\label{subsec:related_works_LLM}
Recent literature has extensively investigated alignment strategies for mapping wireless data into high-dimensional latent spaces compatible with pre-trained architectures \cite{liang2026large}. Since standard LLMs operate on tokenized sequences, the dominant strategy involves projecting wireless features into embeddings that emulate discrete linguistic tokens. A representative framework is LLM4CP \cite{Liu2024channel}, which addresses channel prediction by normalizing and patching CSI sequences in the frequency-delay domain. These patches are subsequently projected via a trainable neural network (NN) embedding layer into token vectors, allowing the LLM to process wireless data as pseudo-language sequences. In contrast, BP-LLM \cite{sheng2025beam} introduces a dual-input paradigm: Natural language prompts are employed to trigger the model’s reasoning capability, while historical beam and angle-of-departure (AoD) sequences are converted into token-like embeddings via a trainable tokenizer. This mechanism effectively reformulates beam prediction as a next-token generation task.
Beyond single-task adaptations, recent works have explored multimodal and multi-task integration using similar tokenization principles. M2BeamLLM \cite{zheng2025m2beamllm} proposes a unified encoding framework where heterogeneous data from sensors (e.g., LiDAR, GPS) and wireless links are fused and aligned into a single token sequence. Similarly, for multi-task physical-layer networks, \cite{zheng2025large} develops specialized LLM embedders that facilitate the simultaneous execution of precoding, signal detection, and channel prediction within a single pre-trained architecture.

Concurrently, an alternative paradigm exploits the structural correlations between wireless channels and natural images to leverage pre-trained large vision models (LVMs). Unlike token-based methods that require learnable embedders, LVM4CSI \cite{guo2025lvm4csi} proposes a direct visualization strategy. By mapping numerical CSI matrices into RGB-compatible formats—either through colormaps or channel-wise mapping of real and imaginary parts—this approach enables the direct application of standard vision backbones (e.g., ConvNeXt) without extensive fine-tuning. Similarly, to address delay-Doppler estimation in mmWave systems, 2DLAM \cite{xie20252dlam} employs a feature extraction module that converts delay-Doppler profiles into image-like embeddings compatible with ImageGPT. These visual alignment strategies effectively circumvent the need for complex tokenization, leveraging the inherent spatial pattern recognition capabilities of LVMs for wireless tasks.

Despite the extensive exploration of alignment strategies—ranging from token-based embeddings to visual transformations—these methodologies are predominantly tailored for physical-layer tasks. In contrast, MAC-layer optimization depends critically on interference topology, which is a relational structure more naturally captured by graph-based models. How to effectively embed this non-Euclidean topology into the sequential reasoning architectures of LLMs remains an open problem. 


\section{System Model and Problem Formulation}\label{sec:system_model}

We consider the power allocation problem in a wireless network comprising a set of $K$ transmitter-receiver pairs that share a common frequency band. The network operates under a quasi-static block fading assumption, where the topology and channel gains remain invariant within a scheduling slot but vary independently across slots. Accordingly, this work focuses on snapshot-based resource optimization for given channel realizations.

Let $h_{kj} \in \mathbb{C}$ denote the complex channel coefficient from transmitter $j$ to receiver $k$. We distinguish between the direct link gain $h_{kk}$ and the cross-link interference channel $h_{kj}$ (for $j \neq k$). The transmit power of the $k$-th transmitter is denoted by $p_k$, subject to a uniform maximum power budget $P_{\max}$. The global power allocation vector is defined as $\boldsymbol{p} = [p_1, \dots, p_K]^\top$. Treating interference from concurrent transmissions as additive noise, the signal-to-interference-plus-noise ratio (SINR) at receiver $k$ is given by
\begin{equation}
\text{SINR}_k(\boldsymbol{p}) = \frac{|h_{kk}|^2 p_k}{\sum_{j \neq k} |h_{kj}|^2 p_j + \sigma^2},
\end{equation}
where $\sigma^2$ represents the noise power. The Shannon capacity for user $k$ is given by
\begin{equation}
R_k(\boldsymbol{p}) = \log_2(1 + \text{SINR}_k(\boldsymbol{p})).
\end{equation}

The primary objective is to maximize a network-wide utility function subject to individual power constraints. Following the formulation in \cite{chowdhury2021unfolding}, the optimization problem is cast as
\begin{equation}
\label{eq:generic_prob}
    \begin{aligned}
    \underset{\boldsymbol{p}}{\text{maximize}} \quad & \sum_{k=1}^{K} \beta(R_k(\boldsymbol{p})) \\
    \text{subject to} \quad & 0 \leq p_k \leq P_{\max}, \quad \forall k \in \{1, \dots, K\},
    \end{aligned}
\end{equation}
where $\beta(\cdot)$ is a strictly increasing, concave utility function. Adopting different forms of $\beta(\cdot)$, we target three distinct objectives:

\begin{itemize}
    \item \textbf{Sum-Rate Maximization:} Setting $\beta(z) = z$ maximizes the total network throughput. However, this objective inherently favors users with strong channel conditions, which frequently leads to the starvation of weak users.
    
    \item \textbf{Proportional Fairness:} Setting $\beta(z) = \log(z)$ is equivalent to maximizing the geometric mean of user rates. This formulation facilitates a balanced trade-off between aggregate throughput and individual user fairness.
    
    \item \textbf{Harmonic Maximization:} Setting $\beta(z) = -z^{-1}$ imposes severe penalties on low data rates. This metric enforces strict fairness and compels the system to prioritize the most disadvantaged links in the network.
\end{itemize}

The optimization problem in \eqref{eq:generic_prob} is non-convex and NP-hard due to the interference coupling in the SINR term. Classical approaches, most notably the WMMSE algorithm \cite{christensen2008wmmse}, address this challenge by iteratively optimizing a convex surrogate. However, such iterative processing incurs prohibitive computational complexity in practice, making it hard to scale to large-size wireless networks. Additionally, owing to the non-convex nature of the objective, algorithms like WMMSE are highly sensitive to initialization. They frequently become trapped in sub-optimal local minima unless computationally expensive multiple random restarts are employed. Furthermore, their practical efficacy deteriorates significantly under strict fairness constraints. Specifically, the WMMSE update rules rely on auxiliary weights derived from the derivative of the utility function. For the harmonic utility $\beta(z) = -z^{-1}$, the derivative scales as $z^{-2}$. In high-interference regimes where the rates $R_k$ of disadvantaged users can become very small, these weights exhibit hyperbolic growth. Such numerical hypersensitivity induces violent oscillations in the iterative updates, leading to severe algorithmic instability.

\begin{figure}[t]
    \centering
    \includegraphics[width=1.0\columnwidth]{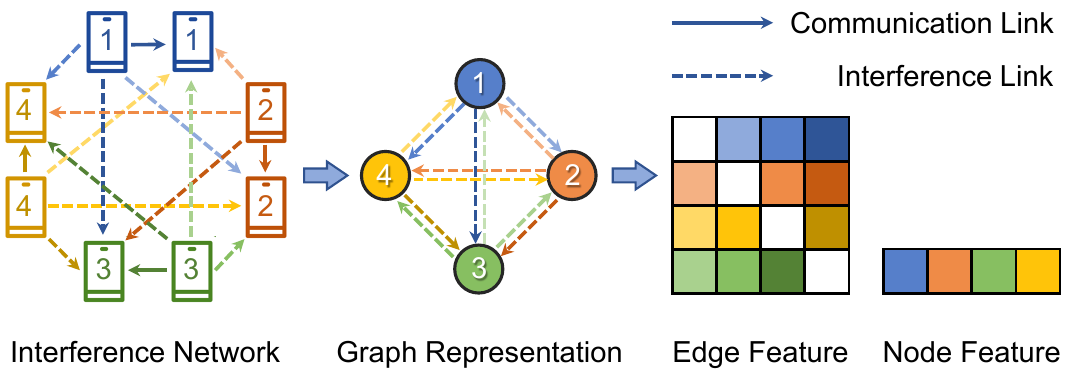}
    \caption{Graph representation of the wireless interference network. Communication links within transmitter-receiver pairs are modeled as nodes, while the mutual interference between pairs is represented by directed edges, constituting a fully connected interference graph $\mathcal{G}$.}
    \label{fig:graph_def}
\end{figure}

To circumvent these numerical pathologies, we propose a paradigm shift toward learning-based approaches that bypass iterative instability. Following the graph representation established in \cite{shen2020graph}, we model the interference topology as a fully-connected directed graph $\mathcal{G} = (\mathcal{V}, \mathcal{E})$, as illustrated in Fig.~\ref{fig:graph_def}. Specifically, each node $v_k \in \mathcal{V}$ represents the direct communication link within transmitter-receiver pair $k$, while each directed edge $e_{kj} \in \mathcal{E}$ characterizes the interference link from transmitter $j$ to receiver $k$. This graph-theoretic formulation explicitly captures the interference patterns that govern system utility, providing a structured topological foundation for the proposed PC-LLM framework.

\section{PC-LLM Framework}

\begin{figure*}[t]
    \centering
    \includegraphics[width=0.72\textwidth]{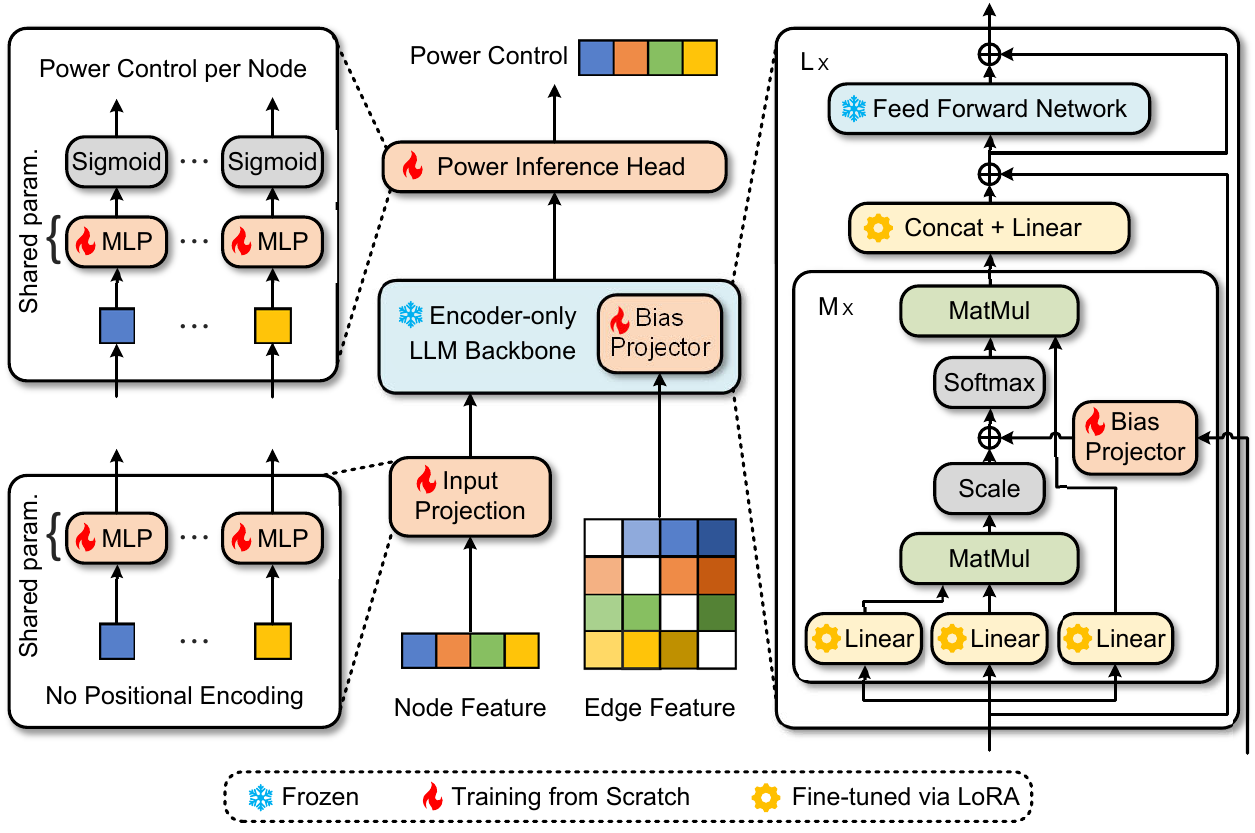}
    \caption{Overall architecture of the proposed PC-LLM. The framework aligns physical channel features with the latent space of the model through an input projection layer, a bias-modulated Transformer backbone, and a power inference head.}
    \label{fig:architecture}
\end{figure*}

This section details the architecture of PC-LLM, a physics-informed framework tailored to repurpose pre-trained LLMs for wireless power control. As illustrated in Fig.~\ref{fig:architecture}, our approach diverges from the local filtering paradigm of standard MPNNs, adopting instead a global relational modeling strategy. While conventional MPNNs encounter severe capacity bottlenecks due to the indiscriminate compression of dense neighborhoods, our framework leverages the global self-attention mechanism to model complex all-to-all interference interactions. By maintaining a pairwise attention matrix, the model avoids the node-level aggregation bottleneck, allowing for selective processing of critical interference paths without lossy compression.

The framework is organized into a three-stage pipeline: 1) \textit{Feature Construction}, which projects raw CSI into structured graph attributes; 2) \textit{Model Architecture}, comprising the embedding layers and the bias-modulated Transformer backbone; and 3) \textit{Training Strategy}, which outlines the parameter-efficient adaptation process.

\subsection{Feature Construction}

To bridge the modality gap between high-dynamic-range wireless signals and the stable latent space of LLMs, we implement a unified preprocessing pipeline. Raw channel gains $|h|^2$ are converted to the decibel (dB) scale and subsequently normalized to achieve zero mean and unit variance. We denote this element-wise transformation as $\psi(|h|^2) = 10\log_{10}(|h|^2)$. Based on this protocol, we define the node and edge features for the interference graph $\mathcal{G}$.

\textbf{Node Feature.} Associated with each node $v_k \in \mathcal{V}$, the feature scalar $s_k \in \mathbb{R}$ encodes the intrinsic physical state of transmitter-receiver pair $k$. It is defined as the normalized direct link strength $s_k = \psi(|h_{kk}|^2)$.

\textbf{Edge Feature.} Associated with each directed edge $e_{kj} \in \mathcal{E}$, the feature vector $\mathbf{z}_{kj}$ characterizes the mutual interference coupling between transmitter-receiver pairs $j$ and $k$. While the graph edge $e_{kj}$ structurally denotes the directional interference from transmitter $j$ to receiver $k$, optimal power allocation necessitates an analysis of the bilateral interaction between the pair. Adopting the bidirectional modeling strategy from \cite{shen2023graph}, we construct a composite feature that concatenates both the incoming interference strength and the reciprocal outgoing interference strength:
\begin{equation}
\mathbf{z}_{kj} = \left[ \psi(|h_{kj}|^2), \psi(|h_{jk}|^2) \right]^\top \in \mathbb{R}^2.
\end{equation}
In contrast to architectures that treat edges as independent tokens \cite{dwivedi2021generalization}, this design maintains an input sequence length consistent with the node cardinality $|\mathcal{V}| = K$. These edge vectors $\mathbf{z}_{kj}$ function as conditioning inputs for the bias projector, explicitly modulating the attention weights between nodes $v_k$ and $v_j$ during the structural reasoning stage.

\subsection{Model Architecture}
The proposed architecture maps the constructed features to optimal power allocation decisions through a sequence of differentiable transformations.

\textbf{Input Projection.} Since the pre-trained Transformer backbone operates on high-dimensional tokens, the scalar node feature $s_k$ requires projection. We employ a lightweight node encoder, implemented as a two-layer MLP with a Gaussian error linear unit (GELU) activation, to map $s_k$ into the hidden dimension $d_{\text{model}}$. Formally, the initial node embedding $\mathbf{x}_k^{(0)} \in \mathbb{R}^{d_{\text{model}}}$ is computed as
\begin{equation}
    \mathbf{x}_k^{(0)} = \mathbf{W}_2 \cdot \text{GELU}(\mathbf{W}_1 s_k + \mathbf{b}_1) + \mathbf{b}_2,
\end{equation}
where $\mathbf{W}_1 \in \mathbb{R}^{d_{\text{mid}} \times 1}$ and $\mathbf{W}_2 \in \mathbb{R}^{d_{\text{model}} \times d_{\text{mid}}}$ are learnable weight matrices, while $\mathbf{b}_1 \in \mathbb{R}^{d_{\text{mid}}}$ and $\mathbf{b}_2 \in \mathbb{R}^{d_{\text{model}}}$ denote the corresponding bias vectors. These embeddings constitute the input matrix $\mathbf{X}^{(0)} = [\mathbf{x}_1^{(0)}, \dots, \mathbf{x}_K^{(0)}]^\top \in \mathbb{R}^{K \times d_{\text{model}}}$.

\textbf{LLM Backbone.} 
The sequence $\mathbf{X}^{(0)}$ is fed into the pre-trained Transformer backbone, concretely implemented using an encoder-only LLM such as BERT \cite{devlin2019bert}. By treating wireless links as input tokens, we adapt the frozen pre-trained model to the power control task via two concurrent mechanisms applied across all $L$ layers: parameter-efficient fine-tuning via LoRA (detailed in Section~\ref{subsec:training_strategy}) and interference-aware bias injection. 
We specifically utilize bidirectional self-attention, as the optimal strategy for any node is contingent upon the global interference state, rendering causal masking inappropriate.
For the $l$-th layer, the input $\mathbf{X}^{(l-1)}$ is projected into query, key, and value matrices for each attention head $m$, represented by
\begin{equation}
\begin{aligned}
    \mathbf{Q}^{(l,m)} &= \mathbf{X}^{(l-1)}\mathbf{W}_Q^{(l,m)}, \\
    \mathbf{K}^{(l,m)} &= \mathbf{X}^{(l-1)}\mathbf{W}_K^{(l,m)}, \\
    \mathbf{V}^{(l,m)} &= \mathbf{X}^{(l-1)}\mathbf{W}_V^{(l,m)}, \\
\end{aligned}
\end{equation}
where $\mathbf{W}_{\{Q,K,V\}}^{(l,m)} \in \mathbb{R}^{d_{\text{model}} \times d_m}$ denote the projection matrices, which will be fine-tuned via LoRA, $d_m = d_{\text{model}}/M$ represents the dimension of each head, and $M$ denotes the number of attention heads.
To strictly enforce the permutation equivariance required for graph processing, we deliberately eliminate the positional encodings and the tokenizer from the standard LLM architecture. This modification ensures that the reasoning process is driven solely by the physical interference topology encoded in the biases.

\textbf{Interference-Aware Bias Injection.} Standard Transformers rely on content-based attention, where relationships are inferred primarily from node feature similarities. This mechanism suffers from topological insensitivity in wireless contexts, as interaction strength is governed by physical interference channels rather than latent feature dot-products. To address this limitation, we propose an interference-aware bias injection mechanism. Inspired by the pairwise-bias design in the Evoformer of AlphaFold2 \cite{jumper2021highly}, we explicitly map the interference topology into the self-attention calculation. 

We introduce a learnable bias projector $\phi^{(l)}(\cdot)$ for each layer, implemented as a separate MLP with a rectified linear unit (ReLU) activation. It maps the edge feature $\mathbf{z}_{kj} \in \mathbb{R}^2$ to a bias value $B_{kj}^{(l,m)}$ for each head $m$, i.e., $[B_{kj}^{(l,1)}, \dots, B_{kj}^{(l,M)}]^\top = \phi^{(l)}(\mathbf{z}_{kj})$, which is defined as
\begin{equation}
    \phi^{(l)}(\mathbf{z}_{kj}) = \mathbf{W}_{\text{bias},2}^{(l)} \cdot \text{ReLU}(\mathbf{W}_{\text{bias},1}^{(l)} \mathbf{z}_{kj} + \mathbf{b}_{\text{bias},1}^{(l)}) + \mathbf{b}_{\text{bias},2}^{(l)},
\end{equation}
where $\mathbf{W}_{\text{bias},1}^{(l)} \in \mathbb{R}^{d_{\text{proj}} \times 2}$ and $\mathbf{W}_{\text{bias},2}^{(l)} \in \mathbb{R}^{M \times d_{\text{proj}}}$ are learnable weight matrices, while $\mathbf{b}_{\text{bias},1}^{(l)} \in \mathbb{R}^{d_{\text{proj}}}$ and $\mathbf{b}_{\text{bias},2}^{(l)} \in \mathbb{R}^{M}$ denote the corresponding bias vectors. We initialize the output layer of the bias projector to zero so that all bias terms start from zero at the beginning of training. This allows the model to leverage the pre-trained attention distribution initially before gradually incorporating the physical interference constraints. The attention score between node $k$ and node $j$ at head $m$ of layer $l$ is modified as
\begin{equation}
    A_{kj}^{(l,m)} = \frac{(\mathbf{q}_k^{(l,m)})^\top \mathbf{k}_j^{(l,m)}}{\sqrt{d_m}} + B_{kj}^{(l,m)},
\end{equation}
where $\mathbf{q}_k^{(l,m)}$ and $\mathbf{k}_j^{(l,m)}$ denote the query vector and the key vector for node $k$ and node $j$, which correspond to the $k$-th and $j$-th row of matrices $\mathbf{Q}^{(l,m)}$ and $\mathbf{K}^{(l,m)}$, respectively. The contextualized representation for layer $l$ is subsequently computed by concatenating the outputs of $M$ attention heads, followed by a linear projection. Let $\mathbf{H}^{(l,m)}$ denote the output of the $m$-th attention head, calculated as
\begin{equation}
\mathbf{H}^{(l,m)} = \text{Softmax}(\mathbf{A}^{(l,m)})\mathbf{V}^{(l,m)}.
\end{equation}

The combined output of the multi-head attention block, denoted as $\tilde{\mathbf{X}}^{(l)}$, is then formulated as
\begin{equation}
\tilde{\mathbf{X}}^{(l)} = \text{Concat}(\mathbf{H}^{(l,1)}, \dots, \mathbf{H}^{(l,M)}) \mathbf{W}_O^{(l)},
\end{equation}
where $\mathbf{W}_O^{(l)} \in \mathbb{R}^{d_{\text{model}} \times d_{\text{model}}}$ denotes the output projection matrix for layer $l$, which is also adapted via LoRA. Subsequently, $\tilde{\mathbf{X}}^{(l)}$ is processed through a standard residual connection and a feed-forward network (FFN) to yield the output of the $l$-th layer, denoted as $\mathbf{X}^{(l)}$. By defining the bias as a function of physical channel gains, the model can prioritize interference-relevant interactions across layers while retaining the global relational reasoning capability of the pre-trained Transformer.

A critical advantage of this approach lies in its computational efficiency. Since self-attention complexity scales quadratically with sequence length, explicitly tokenizing $\mathcal{O}(K^2)$ edges would result in an intractable complexity of $\mathcal{O}(K^4 d_{\text{model}})$. In contrast, by maintaining a node-only sequence and injecting topology directly into attention scores, our approach preserves the standard $\mathcal{O}(K^2 d_{\text{model}})$ complexity, thereby avoiding sequence length explosion and ensuring scalability to dense wireless networks.

\textbf{Power Inference Head.} Following propagation through $L$ Transformer layers, the final latent representation $\mathbf{x}_k^{(L)}$ encapsulates the local state of the node enriched by the global interference context. To map this high-dimensional vector to a feasible power level, we apply a power inference head followed by a Sigmoid activation $\sigma(\cdot)$ scaled by the power budget, represented as
\begin{equation}
    p_k = P_{\max} \cdot \sigma\left( \mathbf{W}_{\text{out}}^\top \cdot \text{GELU}(\mathbf{W}_{\text{hid}} \mathbf{x}_k^{(L)} + \mathbf{b}_{\text{hid}}) \right),
\end{equation}
where $\mathbf{W}_{\text{hid}} \in \mathbb{R}^{d_{\text{hid}} \times d_{\text{model}}}$ and $\mathbf{W}_{\text{out}} \in \mathbb{R}^{d_{\text{hid}}\times 1}$ constitute the learnable weights, with $d_{\text{hid}}$ denoting the hidden dimension of the inference head.

\subsection{Training Strategy}\label{subsec:training_strategy}
The training process is designed to efficiently adapt the general-purpose LLM backbone to the physical characteristics of wireless networks.

\textbf{Parameter-Efficient Fine-Tuning.} To mitigate the risk of overfitting and catastrophic forgetting while reducing memory overhead, we adopt LoRA \cite{hu2021lora} for parameter-efficient fine-tuning. Specifically, we freeze the entire pre-trained LLM backbone and inject trainable low-rank decomposition matrices exclusively into the self-attention projection layers. 
For a specific pre-trained weight matrix $\mathbf{W}_0 \in \mathbb{R}^{d_{\text{out}} \times d_{\text{in}}}$, the update is formally expressed as
\begin{equation}
    \mathbf{W} = \mathbf{W}_0 + \frac{\alpha}{r} \mathbf{W}_{\text{up}} \mathbf{W}_{\text{down}},
\end{equation}
where $\mathbf{W}_{\text{down}} \in \mathbb{R}^{r \times d_{\text{in}}}$ and $\mathbf{W}_{\text{up}} \in \mathbb{R}^{d_{\text{out}} \times r}$ denote the down-projection and up-projection matrices, respectively, with rank $r \ll \min(d_{\text{in}}, d_{\text{out}})$. The scaling factor $\alpha$ is a constant hyperparameter balancing the pre-trained features and the task-specific adaptation. During training, $\mathbf{W}_0$ remains fixed, and only the adapter matrices $\{\mathbf{W}_{\text{down}}, \mathbf{W}_{\text{up}}\}$ are updated. This LoRA adaptation operates jointly with the aforementioned bias injection to align the language model with the wireless interference topologies.

\textbf{Discriminative Learning Rates.} We address the knowledge gap between the pre-trained backbone and the newly initialized components by applying discriminative learning rates \cite{howard2018universal}. Specifically, a higher learning rate is assigned to the node encoder, bias projector, and inference head to encourage rapid adaptation to physical features. Conversely, a lower learning rate is applied to the LoRA parameters to subtly refine the latent space of the backbone, ensuring that the pre-trained relational reasoning capabilities are not disrupted by high-variance gradients.

\textbf{Loss Function.} To enable the discovery of strategies that surpass suboptimal traditional algorithms, we adopt an unsupervised training objective. PC-LLM is trained to directly maximize the system utility defined in Section~\ref{sec:system_model}. The loss function is defined as
\begin{equation}
\mathcal{L} = - \mathbb{E}_{\mathcal{G}} \left[ \sum_{k=1}^{K} \beta\left( R_k(\mathbf{p}) \right) \right],
\end{equation}
where $\mathbf{p}$ is the power vector generated by the power inference head. Minimizing $\mathcal{L}$ is therefore equivalent to maximizing the target utility (e.g., sum-rate, proportional fairness, or harmonic utility) in expectation over the training distribution.

\section{Numerical Results} \label{sec:numerical_results}

\subsection{Experimental Details}

\subsubsection{Simulation Settings}
To evaluate the robustness and generalization capabilities of the proposed framework under realistic deployment conditions, we employ the device-to-device (D2D) underlay network as a representative benchmarking scenario, generating a comprehensive dataset to simulate diverse interference topologies. The simulation environment covers a $1,000 \times 1,000$~m$^2$ square region where $K$ transmitter-receiver pairs are deployed.

\textbf{Topology Generation.} We employ a rigorous topology generation process to mirror practical network configurations. Transmitters are distributed according to a hard-core point process, which enforces a minimum separation distance of $30$~m between nodes to prevent unrealistic physical overlap. Subsequently, a corresponding receiver is generated for each transmitter within a uniform ring area defined by the interval $[d_{\min}, d_{\max}]$. To ensure valid cell association, we enforce a strict nearest-neighbor policy where each receiver must be geographically closest to its serving transmitter. Receiver positions are regenerated if they violate this condition, thereby maintaining valid Voronoi tessellations and guaranteeing that every transmitter serves exactly one user.
The channel model incorporates a dual-slope path loss component combined with log-normal shadowing (standard deviation $\xi = 7$ dB) \cite{naderializadeh2023learning, zhang2015downlink}, alongside Rayleigh fading to capture small-scale variations. 
The system operates over a bandwidth of $W = 10$ MHz, with a maximum transmission power budget of $P_{\max} = 10$ dBm and a noise power spectral density of $-174$ dBm/Hz.

\textbf{Dataset Heterogeneity.} To foster robust generalization across diverse interference environments, we generate a heterogeneous dataset comprising 15 distinct scenarios. These scenarios are formed by the cross-combination of five user density levels $K \in \{20, 35, 50, 65, 80\}$ and three transmitter-receiver distance ranges uniformly sampled from $[d_{\min}, d_{\max}] \in \{ [2, 65], [10, 50], [30, 70] \}$~m. This configuration creates a continuum of interference regimes, ranging from noise-limited sparse networks to interference-limited ultra-dense clusters. Each configuration comprises 70,000 network snapshots, partitioned into 50,000 for training, 10,000 for validation, and 10,000 for testing.

\subsubsection{Performance Metrics} 
Consistent with the formulation in Section~\ref{sec:system_model}, we evaluate the proposed framework using three distinct metrics. Although the training objective maximizes the aggregate network utility $\sum \beta(R_k)$, we report the corresponding mean user rates:
\begin{itemize}
    \item \textbf{Arithmetic Mean Rate:} Evaluates the sum-rate maximization task corresponding to $\beta(z) = z$.
    
    \item \textbf{Geometric Mean Rate:} Evaluates the proportional fairness task corresponding to $\beta(z) = \log(z)$.
    
    \item \textbf{Harmonic Mean Rate:} Evaluates the harmonic maximization task corresponding to $\beta(z) = -z^{-1}$.
\end{itemize}

\subsubsection{Network Hyperparameters and Training Details} 
We employ BERT-Large \cite{devlin2019bert} as the backbone for PC-LLM, which originally comprises approximately 340 million parameters across 24 Transformer layers with a hidden dimension of $d_{\text{model}} = 1,024$ and $M = 16$ attention heads. To balance the performance and inference latency, we utilize the first $12$ Transformer layers (the rationale for this depth selection is provided in Section~\ref{subsec:model_architecture_analysis}), effectively halving the active backbone parameters. We initialize the weights from the standard pre-trained BERT-Large checkpoint, and apply LoRA with a rank $r=16$ and a scaling factor $\alpha=32$ to the query, key, value, and output projection matrices.

\textbf{Optimization Configuration.}  We adopt the discriminative learning rate strategy proposed in Section~\ref{subsec:training_strategy}. The randomly initialized components (node encoder, bias projector, and inference head) are optimized with a learning rate of $\eta_{\text{init}} = 1 \times 10^{-3}$ across all tasks. For the pre-trained backbone, the LoRA learning rate $\eta_{\text{LoRA}}$ is tuned according to the optimization objective: We set $\eta_{\text{LoRA}} = 1 \times 10^{-4}$ for the sum-rate maximization task, and $\eta_{\text{LoRA}} = 3 \times 10^{-4}$ for the proportional fairness and harmonic maximization tasks. The model is trained using the Adam optimizer with a batch size of 512. We utilize an adaptive \texttt{ReduceLROnPlateau} scheduler that decays the learning rate by a factor of 0.5 upon validation stagnation (patience of 15 epochs), allowing the model to thoroughly exploit the optimization landscape.

\textbf{Numerical Stability.} Given the high dynamic range of interference power in ultra-dense networks, numerical stability is critical. We implement a two-tiered safeguard mechanism. First, for the proportional fairness and harmonic objectives, we enforce a lower bound on user rates, $\tilde{R}_k = \max(R_k, \epsilon)$ with $\epsilon=10^{-5}$, to prevent infinite loss values during training. Second, we apply gradient clipping with a maximum norm of 1.0 during backpropagation. This constraint effectively counteracts the exploding gradient problem caused by the steep derivatives of the harmonic utility in low-rate regimes.

\subsubsection{Baselines}

To validate the efficacy of PC-LLM, we compare it against a spectrum of algorithms ranging from classical optimization methods to state-of-the-art learning-based approaches. To ensure a fair comparison regarding model capacity, we specifically introduce scaled-up versions of the deep learning baselines:

\begin{figure*}[ht]
    \centering
    \includegraphics[width=1.0\textwidth]{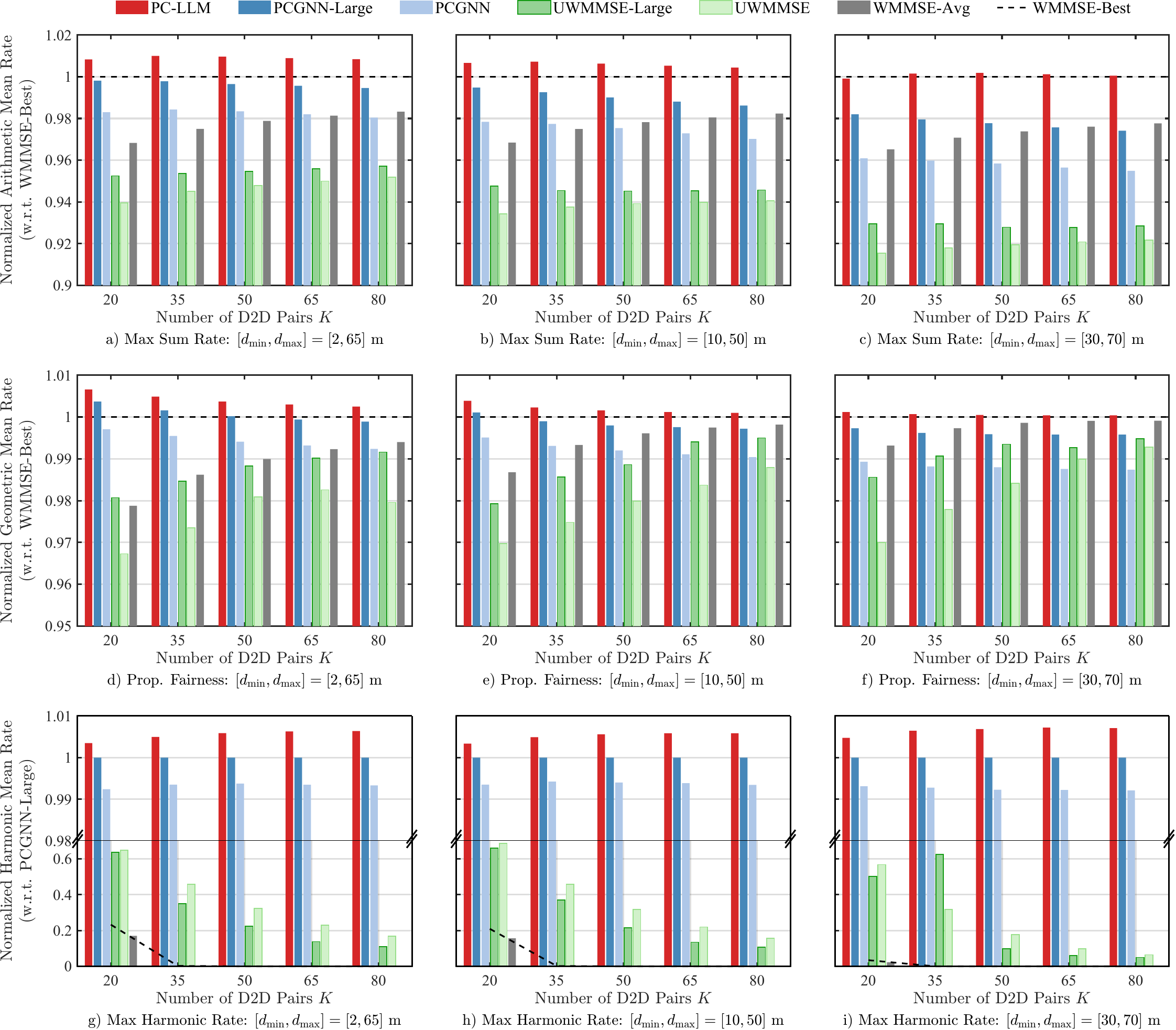}
    \caption{Performance comparison of the proposed PC-LLM framework against baseline algorithms across varying network densities (number of D2D pairs $K$) and interference environments (distance ranges $[d_{\min}, d_{\max}]$). The results are organized by optimization objectives: (a)–(c) sum-rate maximization, (d)–(f) proportional fairness, and (g)–(i) harmonic maximization. The performance is normalized with respect to the WMMSE-Best algorithm for the max sum rate and proportional fairness tasks, whereas the harmonic mean rate is normalized relative to the optimal PCGNN-Large baseline. For learning-based methods, only the best performance achieved under their respective optimal training configurations is reported.}
    \label{fig:main_results}
\end{figure*}

\begin{itemize}
    \item \textbf{WMMSE \cite{christensen2008wmmse}:} The classical WMMSE algorithm serves as the primary optimization benchmark, configured with a maximum of 100 iterations. Given the non-convex nature of the interference management problem, WMMSE is susceptible to initialization-dependent local optima. Therefore, we evaluate two variants: \textbf{WMMSE-Avg} reports the mean utility averaged over 100 independent random initializations, while \textbf{WMMSE-Best} selects the solution yielding the highest utility. Although computationally prohibitive for real-time applications, WMMSE-Best serves as an idealized performance upper bound for iterative optimization methods.
    
        \item \textbf{PCGNN \cite{shen2023graph}:} As a representative MPNN, we evaluate both the standard implementation and a scaled variant, denoted as \textbf{PCGNN-Large}. The latter expands the hidden dimensions of node and edge networks to 1,024 (matching the embedding size of BERT-Large) and thereby increases the message dimension to 3,072. This scaled variant is critical to decouple architectural benefits from parameter counts, verifying that the gains of PC-LLM stem from the proposed bias-tuning mechanism rather than mere model capacity.
    
    \item \textbf{UWMMSE \cite{chowdhury2021unfolding}:} A deep unfolding framework that maps WMMSE iterations into a trainable MPNN architecture. We evaluate both the standard version and \textbf{UWMMSE-Large}, where the hidden dimensions of the internal graph convolutional modules are scaled to 1,024 to align with the capacity of our proposed model.
\end{itemize}

Additionally, we include \textbf{Full Reuse} as a performance lower bound, where all transmitters operate at maximum power $P_{\max}$. This naive strategy is used primarily to quantify the severity of mutual interference and the potential gain offered by power control schemes.


\subsection{Performance Evaluation}\label{subsec:performance_evaluation}

\begin{figure*}[htbp]
\setlength{\abovecaptionskip}{0pt}  
\setlength{\belowcaptionskip}{5pt} 
    \centering
    \begin{minipage}[b]{0.32\textwidth}
        \centering
        \includegraphics[width=\linewidth]{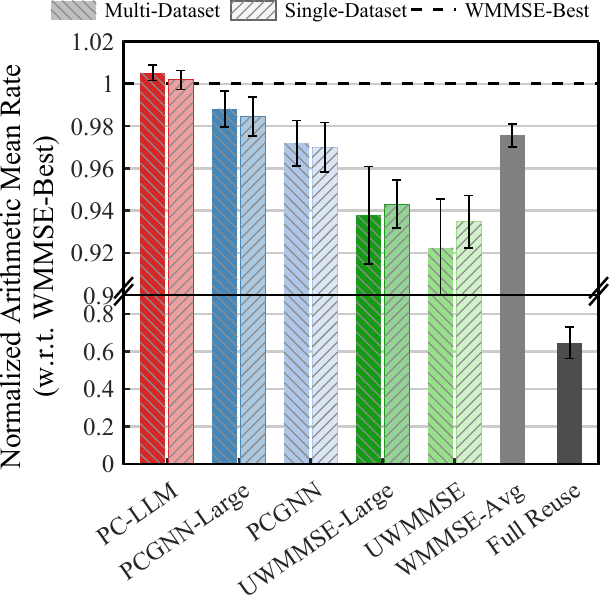}
        \centerline{~~~~~~~~a) Max Sum Rate}
    \end{minipage}
    \begin{minipage}[b]{0.32\textwidth}
        \centering
        \includegraphics[width=\linewidth]{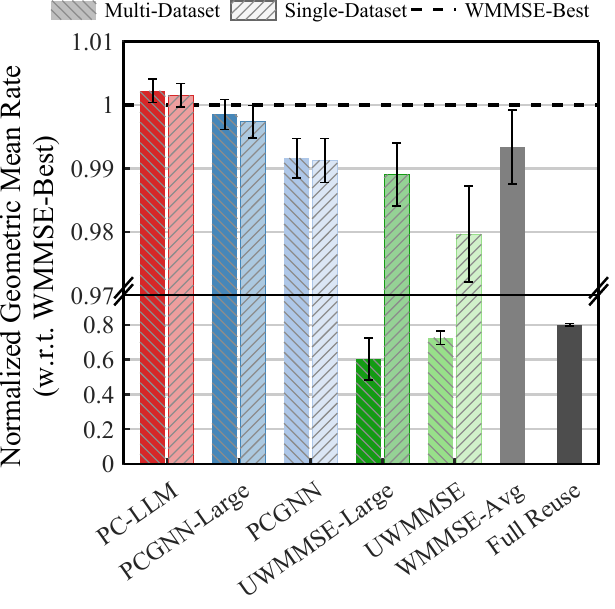}
        \centerline{~~~~~~~~~~b) Prop. Fairness}
    \end{minipage}
    \begin{minipage}[b]{0.33\textwidth}
        \centering
        \includegraphics[width=\linewidth]{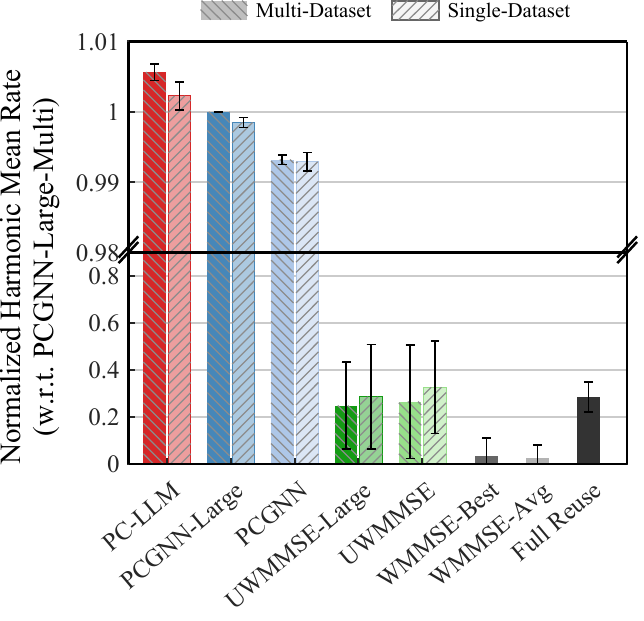}
        \centerline{~~~~~~~~~~c) Max Harmonic Rate}
    \end{minipage}
    \caption{Performance comparison of the proposed PC-LLM and baseline algorithms under multi-dataset (darker bars) and single-dataset (lighter bars) training configurations. The performance is normalized with respect to the WMMSE-Best algorithm for the max sum rate and proportional fairness tasks, whereas the harmonic mean rate is normalized relative to the multi-dataset trained PCGNN-Large baseline. The bar heights represent the average performance across 15 distinct scenarios with varying user densities and channel distributions. The error bars denote the standard deviation, quantifying the performance variance and validating the universal robustness of each method against extreme macro-level topological heterogeneity.}
    \label{fig:single_vs_multi}
\end{figure*}


Fig.~\ref{fig:main_results} benchmarks the performance of the proposed PC-LLM against baseline algorithms across varying network densities (number of D2D pairs $K$) and interference environments (distance ranges $[d_{\min}, d_{\max}]$). The evaluation is stratified by three distinct optimization objectives: sum-rate maximization in (a)--(c), proportional fairness in (d)--(f), and harmonic maximization in (g)--(i). To ensure a rigorous comparison, performance metrics for the first two objectives are normalized with respect to the idealized WMMSE-Best algorithm. While for the harmonic maximization task, results are normalized relative to the optimal PCGNN-Large baseline, an adjustment necessitated by the numerical instability of WMMSE in this regime.
Across the nine cases in Fig.~\ref{fig:main_results}, PC-LLM establishes a consistent dominance, stably surpassing the idealized WMMSE-Best benchmark across varying scales. Notably, the scaled-up baselines, i.e., PCGNN-Large and UWMMSE-Large, consistently outperform their standard counterparts, validating that increased model capacity contributes to performance gains. However, despite matching the parameter scale, PCGNN-Large still lags behind PC-LLM. This performance gap highlights that the superiority of PC-LLM stems not merely from model size, but fundamentally from the structural advantage of the proposed bias-tuned Transformer architecture over standard message-passing paradigms.

Specific to the sum-rate and proportional fairness tasks shown in Fig.~\ref{fig:main_results}(a)--\ref{fig:main_results}(f), most learning-based methods achieve competitive performance in sparse network configurations ($K=20$). However, as the network transitions into interference-limited regimes ($K=80$), the interference landscape becomes substantially more complex. In these high-density scenarios, MPNN-based baselines struggle to maintain parity with the WMMSE-Best benchmark. Resonating with our discussion in Section~\ref{subsec:related_works_GNN}, this limitation is attributed to theinherent representational bottleneck of the aggregation mechanism in MPNNs. As network density escalates, squashing an increasingly large volume of incoming messages into a single fixed-size vector causes severe signal dilution, which obscures dominant interferers. PC-LLM avoids  this lossy compression by explicitly computing an all-to-all interaction matrix before feature fusion, preserving individual interference paths. Consequently, PC-LLM consistently surpasses the idealized WMMSE-Best benchmark across all tested topologies, sustaining a robust outperformance even under the most severe interference conditions. This indicates that the proposed interference-aware bias tuning mechanism effectively guides the Transformer to explicitly model critical interference paths, maintaining stable optimization efficacy across diverse interference densities.

The performance gap is more pronounced in the harmonic maximization task shown in Fig.~\ref{fig:main_results}(g)--\ref{fig:main_results}(i), which imposes stringent fairness constraints. In this regime, we observe a sharp dichotomy between learning-based and optimization-based paradigms. While learning-based methods demonstrate stability, with PC-LLM consistently outperforming the robust PCGNN-Large baseline, optimization-based approaches suffer drastic degradation. Corroborating our analysis in Section~\ref{sec:system_model}, the collapse of WMMSE is attributed to numerical hypersensitivity: The derivative of the harmonic utility scales inversely with the square of the rate ($z^{-2}$), causing auxiliary weights to exhibit hyperbolic growth when edge user rates approach zero. These violent oscillations render the iterative process unstable. Although UWMMSE employs a trainable unfolding network that partially mitigates this instability, its performance is far below other learning-based schemes. In contrast, PC-LLM, as an end-to-end architecture, bypasses gradient-based auxiliary weight calculations, demonstrating exceptional numerical stability and effectively protecting disadvantaged users.


\begin{table*}[htbp]\label{tab:generalization_performance}
    \renewcommand{\arraystretch}{1.2}
    \centering
    \caption{Generalization Performance Across Varying Network Densities with $[d_{\min}, d_{\max}] = [1, 100]$ m}
    \label{tab:generalization_performance}
    \begin{tabular}{@{} c l c c c c c @{}} 
        \toprule
        \multirow{2.5}{*}{\textbf{Optimization Objective}} & 
        \multicolumn{1}{c}{\multirow{2.5}{*}{\textbf{Algorithm}}} & 
        \multicolumn{5}{c}{\textbf{Number of D2D Pairs $K$}} \\
        \cmidrule(lr){3-7} 
        & & \textbf{20} & \textbf{35} & \textbf{50} & \textbf{65} & \textbf{80} \\
        \midrule
        \multirow{3}{*}{Max Sum Rate} 
        & PC-LLM (zero-shot) & \textbf{\underline{4.7741}} & \textbf{\underline{3.8514}} & \textbf{\underline{3.3893}} & \textbf{\underline{3.0993}} & \textbf{\underline{2.9070}} \\
        & PC-LLM (full-shot) & 4.7731 & 3.8183 & 3.3760 & 3.0914 & 2.9011 \\
        & WMMSE-Best         & 4.7452 & 3.8152 & 3.3561 & 3.0701 & 2.8811 \\
        \midrule
        \multirow{3}{*}{Prop. Fairness} 
        & PC-LLM (zero-shot) & \textbf{\underline{2.1553}} & \textbf{\underline{1.3818}} & \textbf{\underline{1.0601}} & \textbf{\underline{0.8838}} & \textbf{\underline{0.7776}} \\
        & PC-LLM (full-shot) & 2.1536 &  1.3511 &  1.0482 & 0.8830 & 0.7702 \\
        & WMMSE-Best         & 2.1422 & 1.3748 & 1.0556 & 0.8805 & 0.7750 \\
        \midrule
         \multirow{3}{*}{\shortstack{Max Harmonic Rate \\ ($\times 10^{-2}$)}} 
        & PC-LLM (zero-shot) & \textbf{\underline{128.0428}} &  \textbf{\underline{66.7253}} & \textbf{\underline{45.0104}} &  \textbf{\underline{34.8976}} & \textbf{\underline{29.0176}} \\
        & PC-LLM (full-shot) & 127.0949 & 66.3301 & 44.8225 & 34.7026 & 28.8410 \\
        & WMMSE-Best         & 16.3880 & 0.1511 & 0.0037 & 0.0033 & 0.0031 \\        \bottomrule
    \end{tabular}
    
\end{table*}

Fig.~\ref{fig:single_vs_multi} investigates the impact of training data diversity on the performance of these learning-based schemes. For the single-dataset configuration, each learning-based model is trained exclusively on 50,000 snapshots specific to the evaluated scenario, which is illustrated using the lighter bars in Fig.~\ref{fig:single_vs_multi}. For the multi-dataset configuration, the models are trained by leveraging 750,000 snapshots aggregated across all 15 distinct scenarios, illustrated using the darker bars in Fig.~\ref{fig:single_vs_multi}. For each algorithmic configuration, the bar height represents the normalized rate averaged across all 15 distinct scenarios, while the error bar indicates the standard deviation. A distinct negative transfer phenomenon is observed for UWMMSE, where the multi-dataset model underperforms its single-dataset counterpart. This suggests that unfolding methods tend to overfit to specific channel statistics and lack the flexibility to generalize across diverse environments. In contrast, PC-LLM demonstrates robust positive transfer, effectively leveraging heterogeneous data to refine its policy, which is remarkable considering its base performance is already near-optimal (i.e., outperforming WMMSE-Best). This validates the superior capacity of the Transformer backbone to assimilate diverse topological data. Fig.~\ref{fig:single_vs_multi} also shows that PC-LLM exhibits remarkably narrow error bars despite the vastly different network topologies across the 15 datasets. This confirms that the model successfully avoids overfitting to specific deployment scales. Instead, it learns a universally robust optimization policy that maintains a highly stable relative advantage over the WMMSE-Best baseline regardless of the network configuration.

\subsection{Generalization Analysis}

Table~\ref{tab:generalization_performance} evaluates the zero-shot generalization capability of the proposed framework. We deploy the zero-shot PC-LLM in an unseen, wide-range scenario. While the model was exclusively trained on a heterogeneous mixture of spatial configurations bounded within $[2, 70]$ m, the evaluation extrapolates the transmitter-receiver distances to a broad, unseen range of $[1, 100]$ m. This extrapolation subjects the model to a substantially broader dynamic range of interference power than distribution encountered during the training phase. We benchmark this transferability against two rigorous baselines: the idealized optimization upper bound provided by WMMSE-Best, and PC-LLM (full-shot), which is the proposed framework fine-tuned on the target distribution, following the same single-dataset methodology in Fig~\ref{fig:single_vs_multi}.
Despite the severe distribution shift, PC-LLM demonstrates exceptional robustness. It consistently outperforms the iterative WMMSE-Best benchmark across all network densities and tasks, while achieving performance on par with, and occasionally exceeding, the explicitly trained full-shot model. This performance parity between zero-shot extrapolation and full-shot specialization highlights the structural efficacy of the proposed physics-informed architecture. By explicitly integrating physical channel states into the attention calculation, the interference-aware bias tuning grounds the relational reasoning of the language model in actual wireless physics. Instead of overfitting to the statistical artifacts of specific training topologies, the framework distills a universal optimization policy. Consequently, the learned strategy remains highly robust and near-optimal even when spatial configurations differ significantly from the training scenarios.

\subsection{Model Architecture Analysis}\label{subsec:model_architecture_analysis}

\begin{figure*}[ht]
    \centering
    \includegraphics[width=1.0\textwidth]{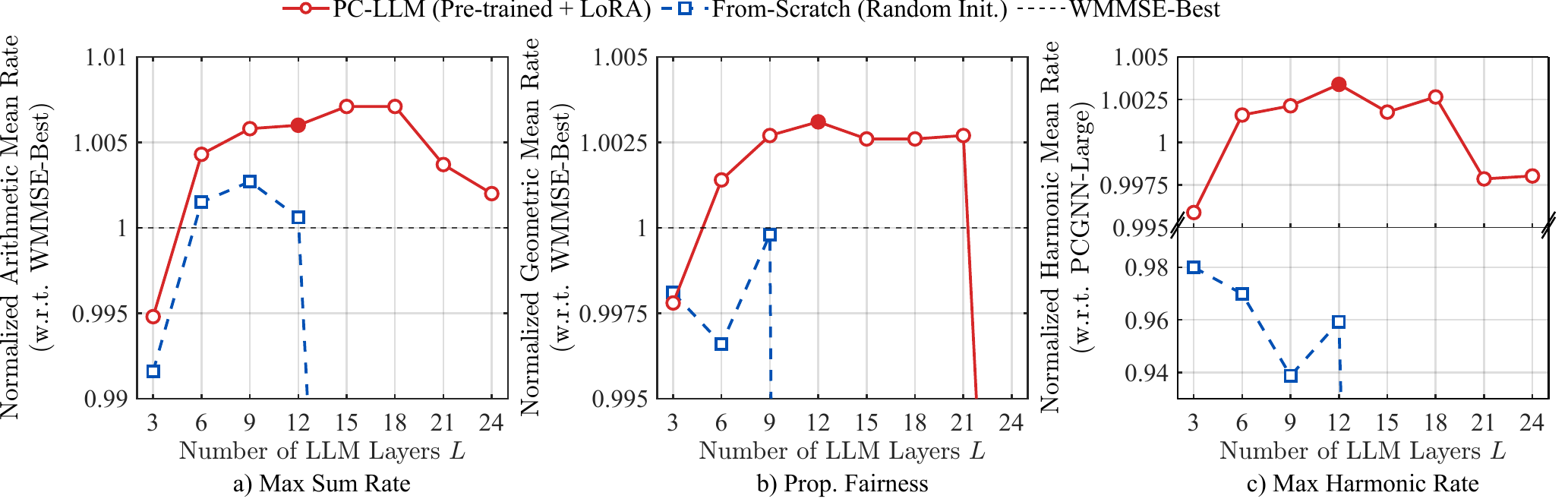}
    \caption{Performance comparison with varying numbers of LLM layers $L$ under the setup of $K=50$ and $[d_{\min}, d_{\max}] = [2, 65]$ m. The performance is normalized with respect to the WMMSE-Best algorithm for the max sum rate and proportional fairness tasks, whereas the harmonic mean rate is normalized relative to PCGNN-Large.}
    \label{fig:model_architecture_analysis}
\end{figure*}

Fig.~\ref{fig:model_architecture_analysis} shows the impact of model depth $L$ on optimization efficacy, evaluated under a representative dense interference environment characterized by $K=50$ pairs and a distance range of $[2, 65]$ m. The results unveil a distinct inverted-U trend. Specifically, in the shallow regime ($L=3$ to $6$), performance improves rapidly, indicating that a reasonable depth is essential to capture the basic interference topology. This upward trend saturates around $L=12$, forming a performance plateau where PC-LLM achieves state-of-the-art results across all objectives. However, as the depth extends further into the deeper regime ($L > 18$), the performance trajectory reverses. A discernible degradation is observed as $L$ approaches 24, particularly in the sum-rate and harmonic maximization tasks, where the normalized rates drop precipitously. We note that early Transformer layers are known to emphasize general relational and dependency patterns, which are more transferable to wireless interference modeling. In our setting, these shallow layers appear to provide the most useful knowledge for understanding and capturing the wireless interference topologies. In contrast, deeper layers in pre-trained LLMs are increasingly specialized in abstract linguistic semantics, such as sentiment or complex reasoning. These semantic features constitute modal noise, creating a modality gap that interferes with wireless power control. Therefore, truncating the model at $L=12$ effectively filters out this noise while retaining the structural reasoning capabilities.

To isolate the contribution of pre-trained knowledge, we benchmark against a From-Scratch baseline. In this configuration, the pre-trained weights are discarded in favor of random initialization, and the LoRA module is removed to permit full-parameter updates using a uniform learning rate $\eta_{\text{init}}$. As depicted in Fig.~\ref{fig:model_architecture_analysis}, a critical comparison reveals a profound performance gap: The randomly initialized counterpart achieves substantially lower performance and exhibits noticeable degradation as network depth increases. This empirical evidence confirms that the superior performance of PC-LLM stems not merely from the Transformer architecture itself but from the rich structural priors embedded in the pre-trained backbone. 
In the absence of these priors, the Transformer remains a generic architecture that lacks inherent structural guidance, failing to organize its attention mechanism for complex physical mappings when trained on limited wireless datasets. In contrast, the pre-trained structural priors provide a robust initialization that effectively guides the model through the highly non-convex optimization landscape of power control. Leveraging these established relational dependencies, PC-LLM successfully captures the underlying interference manifold, whereas the same task remains highly sample-inefficient and fails to reach competitive solutions when attempted from scratch.

\subsection{Ablation Studies}

\begin{figure}[tbp]
    \centering
    \includegraphics[width=0.48\textwidth]{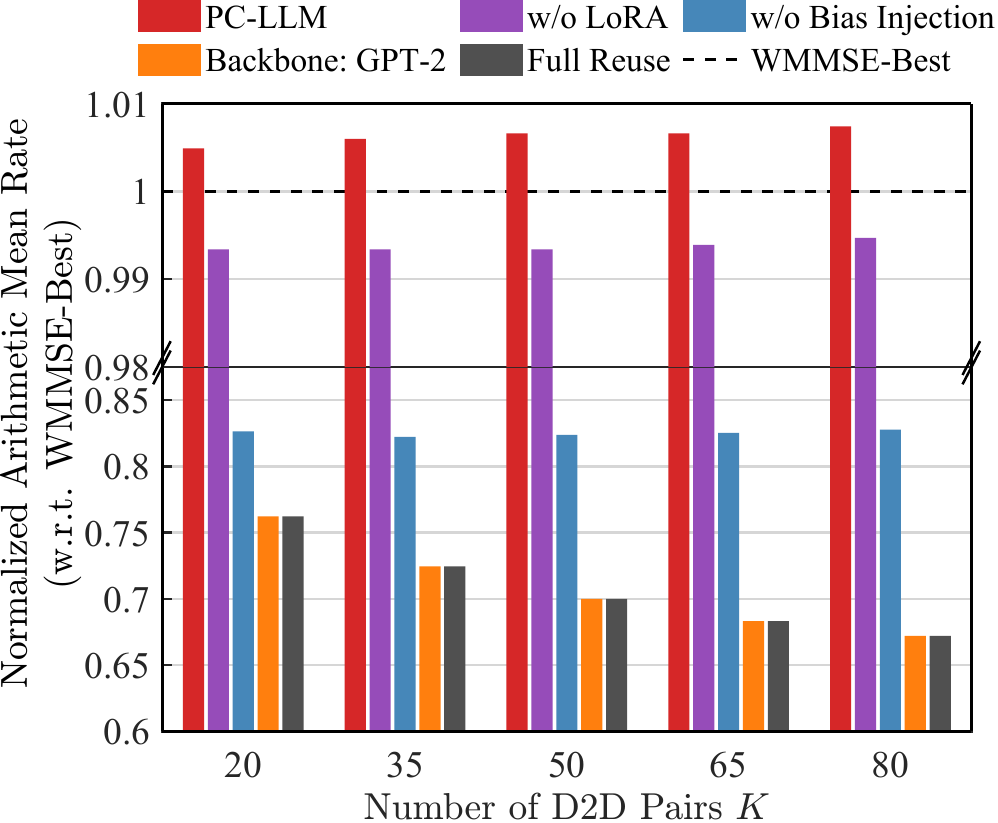}
    \vspace{-3mm}
    \caption{Ablation results of the proposed PC-LLM framework with varying numbers of D2D pairs $K$ under the $[d_{\min}, d_{\max}] = [2, 65]$ m scenario. The performance is normalized with respect to the WMMSE-Best algorithm.}
    \label{fig:ablation}
\end{figure}

Fig.~\ref{fig:ablation} presents an ablation study of the proposed framework to isolate the individual contributions of its core components: the backbone architecture, the interference-aware bias, and the LoRA fine-tuning strategy. The evaluation benchmarks the normalized performance across varying user densities $K$ under the standard interference environment where transmitter-receiver distances are distributed within $[2, 65]$ m.

First, we examine the structural foundation by replacing the bidirectional BERT-Large backbone with a unidirectional GPT-2-Medium (labeled ``Backbone: GPT-2''). This substitution results in a catastrophic performance collapse, with rates degrading to the level of the naive Full Reuse baseline. This failure stems from a fundamental mismatch between the model architecture and the wireless interference management problem. Wireless interference is inherently omni-directional and non-causal, forming a fully connected graph where every node interacts with every other node. In contrast, GPT-2 imposes a causal mask that enforces a strict left-to-right processing order, effectively severing the connectivity of the interference graph. Consequently, the bidirectional attention mechanism of BERT is proven to be not merely a capacity choice, but a structural necessity to preserve the mathematical isomorphism between the attention map and the physical interference topology.

Equally critical is the role of the proposed interference-aware bias. As observed in the ``w/o Bias'' configuration, removing this module causes the normalized rate to plummet precipitously, falling significantly below the WMMSE benchmark. This collapse confirms that standard self-attention is fundamentally topology-blind; without explicit positional or structural encoding, it treats the network as a permutation-invariant set rather than a spatial graph. The proposed bias is therefore decisive, as it injects the physical connectivity matrix directly into the attention mechanism. By modulating attention scores according to physical interference strength, it grounds the model's reasoning in the real-world interference landscape, ensuring that the learned features generalize to wireless power control applications. 

Finally, we analyze the contribution of the training strategy by evaluating the ``w/o LoRA'' configuration, where the pre-trained backbone is frozen and only the bias module and projection heads are updated. Remarkably, the frozen model yields competitive performance near the WMMSE-Best baseline, corroborating our earlier finding that pre-trained weights naturally possess robust structural reasoning capabilities. However, to consistently surpass the state-of-the-art, LoRA is essential for domain adaptation. It bridges the gap between the generic linguistic reasoning of the backbone and the specific distributions of wireless channels, which is essential for network interference management. 

\section{Conclusion}
In this paper, we have developed PC-LLM, a physics-informed framework that repurposes pre-trained LLMs as relational reasoning backbones for MAC-layer power control. To overcome the topological insensitivity of standard Transformers, an interference-aware bias tuning mechanism has been exploited to explicitly inject physical channel states into the LLM attention module. We demonstrate that through this attention-level fusion of wireless topology and pre-trained structural priors, the proposed framework consistently outperforms optimal WMMSE and MPNN baselines with exceptional zero-shot robustness. Our analysis further shows that topology management relies primarily on shallow structural layers rather than deeper semantic ones. This work highlights the potential of wireless graph foundation models, envisioning a future paradigm where models are pre-trained directly on heterogeneous wireless graphs to encode universal physical laws for general-purpose wireless decision-making.


\small
\bibliographystyle{IEEEtran}
\bibliography{journal_dc.bib}

\end{document}